
\input phyzzx

\font\twelverm=cmr10 scaled\magstep1

\font\twelveit=cmti10 scaled\magstep1
\def\twelveb{\bf}
\def\ctrline{\centerline}
\def\r{\rangle}
 \noindent
 To be published in {\it Found. Phys.} July.

\noindent
TAUP 1865-91, hep-th/9305002
\vskip 2 true cm
\twelverm \ctrline{\twelveb{QUANTUM MECHANICAL
INTERACTION-FREE MEASUREMENTS }}
\bigskip
\bigskip
\ctrline{Avshalom C.
Elitzur$^{(a)}$
and Lev Vaidman}

\ctrline{\tenrm{School of
Physics and Astronomy}}
\ctrline{\tenrm{Tel-Aviv University}}
\ctrline{\tenrm{69 978 \ Tel Aviv, ISRAEL}}
\smallskip
\bigskip
\bigskip
\bigskip
\bigskip
\bigskip
\bigskip
\bigskip
\bigskip

\baselineskip 22pt

A novel manifestation of nonlocality of quantum mechanics is
presented.  It is shown that it is possible to ascertain the existence
of an object in a given region of space without interacting with it.
The method might have practical applications for delicate quantum
experiments.
\bigskip
\bigskip
\bigskip
\bigskip

PACS numbers: 03.65.Bz, 42.10.Jd, 42.50.Dv, 42.50.Wm
\eject

\vskip 0.6cm
{\twelveb 1. INTRODUCTION}

Nonlocality is an intriguing aspect of quantum mechanics.  Bell's
inequality$^1$ showed that nonlocality must exist, and Aspect$^2$
provided an experimental proof.  We shall present here yet
another manifestation of the nonlocality of quantum mechanics.  We
shall describe a measurement which, when successful, is capable of
ascertaining the existence of an object in a given region of space,
though no particle and no light ``touched" this object.  This is a new
type of an interaction-free quantum measurement which has no classical
analog.

Let us begin with a brief review of nonlocal measurements which
yield information about the existence of an object in a given region
of space.

If an object is charged or has an electric (magnetic) moments, then
its existence in a given region can be inferred without any particle
passing through that region, but rather by the measurement of the
electric (magnetic) field the object creates outside the region.
Quantum mechanics allows inferring the existence of an object in a
nonlocal way via Aharonov-Bohm effect$^3$ even when the object creates
no electromagnetic field outside a certain space region, but only an
electromagnetic potential.

Even if the object creates no detectable change at a distance, i.e.,
it interacts with the external world only locally, its location can
often be found in a simple nonlocal interaction-free measurement
(i.e., without interacting with the object).  For example, assume it
is known that an object is located in one out of two boxes.  Looking
and {\twelveit not} finding it in one box tells us that the object is
located inside the other box.  A more sophisticated example of
obtaining information in a nonlocal way is the measurement performed
on a system prepared in the Einstein-Podolsky-Rosen state.  If two
objects are prepared in an eigenstate of relative position, the
measurement of the position of one object yields the position of the
other.

In the above cases, what allowed us to infer that an object is
located in a given place by performing an interaction-free measurement
was the information about the object prior to the measurement.  In the
first example we knew that the object is located inside one of the two
boxes, and in the second example we knew about the correlation between
the position of one object and that of another.  The question we
address in this Letter is this: Is it possible to obtain knowledge
about the existence of an object in a certain place using
interaction-free measurements {\twelveit without any prior
information} about the object?  The answer is, indeed, in the
affirmative as we proceed to show.

Our method is based on a particle interferometer which is analogous to
the Mach-Zehnder interferometer of classical optics.  In principle, it
can work with any type of particles.
A particle reaches the first beam splitter which has the
transmission coefficient ${1\over2}$.  The transmitted and reflected
parts of the particle's wave are then reflected by the mirrors in such
a way that they are reunited at another, similar beam splitter
(Fig.~1).  Two detectors collect the particles after they pass through
the second beam splitter.  We can arrange the positions of the
beam splitters and the mirrors so that, due to the destructive
interference, no particles are detected by one of the detectors, say
$D_2$ (but all are detected by $D_1$).  If, without changing the
positions
of the mirrors and the beam splitters, we block one of the two arms of
the interferometer, the particles which succeeded to pass through the
interferometer are detected with equal probability by both detectors
$D_1$ and $D_2$.  Thus, detector $D_2$ detects particles only if
something  stands on the way of particles in one of the routes of the
interferometer.

A practical realization of such an interferometer with electrons and
protons is hampered by strong electromagnetic interaction with the
environment, but neutron interferometers operate in many
laboratories$^4$.  However, our method requires a {\twelveit single
particle interferometer}, i.e. an interferometer with one particle
passing through it at a time, and there is no appropriate neutron
source which produces a single particle states.  Recently$^5$
experiments were performed with a source of single photon states.
Thus we propose to use the Mach-Zehnder interferometer with such a
source of single photons.

\vskip 0.6cm

{\twelveb 2. HOW TO FIND AN OBJECT
WITHOUT INTERACTING WITH IT?}

Our procedure for finding out about the existence of an object in a
given place, without passing even one photon through it, is as
follows: We arrange a photon interferometer as described above, i.e.
no
photons are detected by $D_2$ when both routes of the interferometer
are open, and position it in such a way that one of the routes of the
photon passes through the region of space where we want to detect the
existence of an object (Fig.~1).  We send a single photon through the
system.  There are three possible outcomes of this measurement:
{\obeylines \parskip 0pt
i)~no detector clicks,
ii)~detector $D_1$ clicks,
iii)~detector $D_2$ clicks.}
\noindent
In the first case, the photon has been absorbed (or scattered) by the
object and never reached the detectors.  The probability for this
outcome is ${1\over2}$.  In the second case (the probability for which
is $1\over4$) , the measurement has not succeeded either.  The photon
could have reached $D_1$ in both cases: when the object is, and when
the object is not located in one of the arms of the interferometer.
In this case there has been no interaction with the object so we can
try again.  Finally, in the third case, when the detector $D_2$ clicks
(the probability for which is ${1\over4}$), we have achieved our goal:
we know that there is an object inside the interferometer without
having ``touched" the object.  Indeed, we saw that the necessary
condition for $D_2$ to detect a photon is that one of the routes of
the interferometer is obstructed; therefore the object must be
there.
This is an interaction-free measurement because we had only one photon
and has it interacted with the object, it could never reach detector
$D_2$.$^6$

The quantum mechanical formalism describing the operation of our
device is simple. Let us designate the state of
the photon moving to the right by $\vert 1 \r$, and the state of the
photon
moving up by $\vert 2 \r$.  In a model which illustrates
the essential aspects of the procedure
every time a photon is reflected the phase of its wave
function changes by $\pi\over2$.  Thus, the operation of the
half-silvered plate on the state of the photon is
 $$\eqalign{\vert 1 \r
{}~\rightarrow &~ {{1}\over{\sqrt{2}}}\lbrack\vert 1\r+i\vert 2 \r\rbrack\
,\cr \vert 2 \r ~\rightarrow &~{{1}\over{\sqrt{2}}}\lbrack\vert 2 \r+i\vert 1
\r\rbrack .\cr}\eqno(1)
 $$ The operations  of the two fully-silvered
mirrors are described by $$\vert 1 \r~\rightarrow~ i\vert 2
\r,\eqno(2a) $$
and $$\vert 2 \r~\rightarrow~ i\vert 1 \r .\eqno(2b) $$

If the object is absent, i.e. we have a standard (undisturbed) photon
interferometer, the evolution of the photon's state is described by:
$$\vert 1 \r~\rightarrow ~{{1}\over{\sqrt{2}}}\lbrack\vert 1 \r+i\vert
2 \r\rbrack ~\rightarrow~ {{1}\over{\sqrt{2}}}\lbrack i\vert 2
\r-\vert 1 \r\rbrack ~\rightarrow~ {{1}\over{2}}\lbrack i\vert 2
\r-\vert 1 \r\rbrack - {{1}\over{2}}\lbrack \vert 1 \r+i\vert 2
\r\rbrack= -\vert 1\r.\eqno(3)$$
The photon, therefore, leaves the interferometer moving to the right
towards detector $D_1$, which then clicks.  If, however, the object is
present, the evolution is described by:
$$\vert 1 \r\rightarrow {{1}\over{\sqrt{2}}}\lbrack\vert 1
\r+i\vert 2 \r\rbrack \rightarrow {{1}\over{\sqrt{2}}}\lbrack i\vert 2
\r+i\vert scattered \r\rbrack \rightarrow {{1}\over{2}}\lbrack i\vert
2 \r-\vert 1 \r\rbrack + {{i}\over{\sqrt{2}}}\vert scattered
\r,\eqno(4) $$
where $\vert scattered \r$ is the state of the photon scattered by the
object.  According to the standard approach to quantum
measurement,$^7$ the detectors cause the collapse of the quantum state
(4):
 $${{1}\over{2}}\lbrack i\vert 2 \r-\vert 1 \r\rbrack +
{{i}\over{\sqrt{2}}}\vert scattered \r ~ \rightarrow~ \cases { \vert 2
\r,~~~~ &$D_2$ clicks,~~~ probability~ $1\over4$,\cr \vert 1 \r,~~~~
&$D_1$ clicks,~~~ probability~ $1\over4$,\cr \vert scattered \r~~ &no
clicks,~~~ probability~ $1\over2$.\cr} \eqno(5) $$
We see that the photon can be detected by detector $D_2$ only if the
object is present.  Thus, the click of the detector $D_2$ yields the
desired information, namely, that the object is located somewhere
along the arms the interferometer.
 If we wish to specify by the interaction-free procedure an exact
position of the object inside the interferometer, we can test
(locally) that all but that region inside the interferometer is empty.

The information about the existence of the object was obtained without
``touching" it.  Indeed, we had a single photon.  Had it been
scattered or absorbed (i.e. ``touched") by the object it would not be
detected by $D_2$. Our procedure is, therefore, an interaction-free
measurement of the existence of the object.

The argument which claims that this is an interaction-free measurement
sounds very persuasive, but is, in fact, an artifact of a certain
interpretation of quantum mechanics. (The interpretation which is
usually adopted in discussions of Wheeler's delayed-choice
experiment.) The paradox of obtaining
information without interaction appears due to the assumption that
only one ``branch" of a quantum state exists.  This paradox can be
avoided in the framework of the Many-Worlds Interpretation$^8$ (MWI)
which, however, has paradoxical features of its own.
In the MWI there is no collapse and all ``branches"
of the photon's state (5) are real.  These three branches correspond
to three different ``worlds".  In one world the photon is scattered by
the object and in two others it does not.  Since all worlds take place
in the physical universe we cannot say that nothing has ``touched" the
object.  We get information about the object without touching it in
one world but we ``pay" the price of interacting with the object
in the other world.

\vskip 0.6cm
{\twelveb 3. INTERACTION-FREE COLLAPSE OF A QUANTUM
STATE}

We use here the term ``interaction-free measurement" following
Dicke$^9$. Simplifying an example presented in his paper, we may
consider replacement the object of the previous discussion by a
particle being in a superposition of two states
 $${\vert \Psi \r = \alpha \vert A \r + \beta \vert B \r},\eqno(6)$$
where $\vert A \r$ is a state in which the particle is located in a
small region of space A, and $\vert B \r$ is a state in which the
particle is located in a disjoint small region B. Looking (via
photons) and {\twelveit not finding} a particle in the region A is an
interaction-free measurement of the existence of the particle in B.
The photons passing through the region A are neither scattered nor
absorbed by the particle, therefore there is no interaction with the
particle in this measurement.

One might think in a loose way that the photons passing through the
region A ``push" the wave function of the particle out of region A
causing its collapse into region B. Consider, however, a modification
of this experiment using our procedure for {\twelveit finding} the
particle inside region A without any photon interacting with it.
We place the interferometer in such a way that one of the photon's
routes passes through region A. We send a photon through the
interferometer.  Then, the quantum state of the photon and the
particle becomes:
$$\vert 1 \r\
\vert \Psi \r
{}~\rightarrow~
\alpha\lbrack {{1}\over{2}}\lbrack i\vert 2 \r-\vert 1 \r\rbrack +
{{i}\over{\sqrt{2}}}\vert scattered \r\rbrack \vert A \r +
\beta \vert 1 \r\vert B \r .
\eqno(7)$$
Assuming that detectors cause the collapse of the quantum state, the
evolution of the state (the right hand side of Eq.(7)) continues:
$$ {\rm R.H.S.~of}~ (7)
{}~\rightarrow~
\cases
 { \vert 1\r~\lbrack {{{2}\beta}\over{\sqrt{\alpha^2 + 4\beta^2}}}
 \vert B \r -
{{\alpha}\over{\sqrt{\alpha^2 + 4\beta^2}}}
\vert A \r \rbrack,~~~~ &$D_1$ clicks,~~~ probability
${{\alpha^2}\over{4}} + \beta^2 $,\cr
\vert 2\r~ \vert A\r ,~~~~ &$D_2$ clicks,~~~ probability ${{
\alpha^2}\over{4}},$
\cr
 \vert scattered \r ~\vert A \r~ &no clicks,~~~ probability~
${{\alpha^2}\over{2}}$.\cr } \eqno(8) $$
We see that in the case that detector $D_2$ clicks, the quantum state
of the particle collapses into the state $\vert A \r$ and
 the photon does not ``touch" the particle.  What we have
here is an {\twelveit interaction-free collapse} of the quantum state
of a particle in the box, not only when the particle is not
there, but even when it {\twelveit is} there.$^{10}$

\vskip 0.6cm

{\twelveb 4. HOW TO TEST A BOMB
WITHOUT EXPLODING IT?}

The idea of our Letter is most dramatically illustrated in a way which
is free from any specific interpretation of quantum theory and any
specific meaning of the words ``interaction-free", ``without
touching", etc.  Consider a stock of bombs with a sensor of a new
type: if a single photon hits the sensor, the bomb explodes.  Suppose
further that some of the bombs in the stock are out of order: a small
part of their sensor is missing so that photons pass through the
sensor's hole without being affected in any way, and the bomb does not
explode.  Is it possible to find out bombs which are still in order?

Of course, we can direct some light at each bomb.  If it does not
explode it is not good.  If it does, it {\twelveit was} good.  But we
are interested in finding a good bomb without destroying it.  The
trouble is that the bomb is designed in such a way that {\twelveit
any} interaction with light, even a very soft photon bouncing on
bomb's sensor, causes an explosion.  The task therefore seems to be
impossible, and in classical physics it surely is.  However, our
interaction-free quantum measurement yields a solution.

We place a bomb in such a way that its sensor is located in one of the
possible routes of the photon inside the interferometer.  We send
photons one by one through the interferometer until either the bomb
explodes or detector $D_2$ detects the photon.  If neither of the
above happens, we stop the experiment after a large number of
photons have passed the interferometer.  In the latter case we can
conclude that this given bomb is not good, and we shall try another
one.  If the bomb is good and exploded, we shall also start all over
again with the next bomb.  If, however, $D_2$ clicks, then we achieved
what we promised: we know that this bomb is good and we did not
explode it.  Let us see what is the probability for such an outcome.

If the bomb is good, then for the first photon there is the
probability of $1\over2$ to explode the bomb and $1\over2$ to reach
the second half-silvered mirror through the second route.  The
photon
which reaches the second half-silvered mirror has equal probability to
be detected by $D_1$ and by $D_2$.  In case the photon is detected by
$D_2$ we know that the bomb is good and has not being destroyed.  If
$D_1$ detects the photon we get no information about the bomb, it
might be good or bad.  (In fact the detector $D_1$ can be removed.)
Thus, if the bomb is good, we have the following probabilities:
$1\over4$ to learn that it is good without destroying it, $1\over2$ to
explode it, and $1\over4$ to leave it without getting decisive
information.  In the latter case we should continue and send the next
photon.  The probabilities for the next photon are the same.  Thus,
the good bomb will be found by the second photon with the probability
of ${{1}\over{4}}\cdot {{1}\over{4}}$, by the third photon with the
probability of ${{1}\over{4}}\cdot {{1}\over{4}} \cdot {{1}\over{4}}$
, etc.  The total probability that a good bomb will be found by this
method without being destroyed is, then, $\sum_{n=1}^{\infty}
{{1}\over{4^n}}= {{1}\over{3}} $.

The probability of one third follows from the fact that for each
photon
 $${{{\rm probability~of~hitting~ detector}~D_2}\over{
{\rm probability~of~ hitting~the~bomb}}}~=~ {{1}\over{2}}.$$
This ratio, however, can be improved through small modification of the
procedure.  Let us change the half-silvered plates in the
interferometer such that the first mirror is almost transparent and
the second one is almost not transparent.  The action of the first
beam splitter is described, then, by:
 $$\eqalign{\vert 1 \r
{}~\rightarrow &~ a\vert 1\r+ib\vert 2 \r ,\cr \vert 2 \r ~\rightarrow
&~ a\vert 2 \r+ib\vert 1 \r ;\cr}\eqno(9)$$
and the action of the second beam splitter is:
 $$\eqalign{\vert 1 \r ~\rightarrow &~ b\vert
1\r +ia\vert 2 \r,\cr \vert 2 \r ~\rightarrow &~ b\vert 2 \r+ia\vert 1
\r .\cr}\eqno(10)$$
 where $a$ and $b$ are real and positive, $a^2 + b^2 =
1, a>>b$.  Now, it is easy to see, similarly to Eq.(3), that if the
bomb is not good (in which case it is transparent) then, the photon
starting in the state $\vert 1\r$ ends up in the state $-\vert 1\r$
and, therefore, detector $D_2$ never clicks.  If, however, the bomb is
good, then the evolution of the state of the photon passing through
the interferometer is:
 $$\vert 1 \r\rightarrow a\vert 1 \r+ib\vert 2
\r \rightarrow ia\vert 2 \r+ib\vert absorbed \r \rightarrow ia\lbrack
b\vert 2 \r+ia\vert 1 \r\rbrack + ib\vert absorbed \r.\eqno(11) $$
And, due to the collapse, it continues: $$ {\rm R.H.S.~of}~ (11)
{}~\rightarrow~ \cases { \vert 1\r,~~ &$D_1$ clicks, no explosion~~~
probability $a^4$,\cr \vert 2\r,~~ &$D_2$ clicks, no explosion~~~
probability $a^2b^2,$ \cr \vert absorbed \r, ~~ &no clicks,
explosion~~~~~~~ probability~ $b^2$.\cr } \eqno(12) $$
Thus, $${{{\rm probability~ of~ hitting~ detector}
D_2}\over{ {\rm probability~ of~ hitting~ the~ bomb}}} = a^2.$$

Since $a$ is close to 1, we can test without destroying about a
half out of the good bombs.$^{11}$ Note that in this case the
probability to test the bomb by one photon is small, so we will need
many photons, or we can use the same photon over and over again.

  In one respect the experiment which
tests a bomb without exploding it is easier than the experiment of
testing the existence of an object in a given place without touching
it.  For the latter, in order to ensure that we indeed do not touch
the object, we need a single-particle interferometer.  We could
deduce that no photon scattered by the object because there was only
one photon and had it been scattered by the object it would not be
detected by $D_2$.  For the experiment with the bomb, however, the
source of single particle states is not necessary.  We know that no
photon had touched it simply by the fact that it did not explode.  A
weak enough source, which is stopped once detector
$D_2$
clicks, serves our purpose.  Even the probability of finding a good
bomb remains the same: in the optimal regime about one half of the
good bombs are tested without being destroyed.

\vskip 0.6cm

 {\twelveb 5. CONCLUSIONS}

Our method allows to detect the existence of any unstable system
without disturbing its internal quantum state.  It might, therefore,
have practical applications.  For example, one might select atoms in a
specific excited metastable state.  Let us assume that the atom has
very high crossection for absorbing photons of certain energy while it
is in one out of several metastable states into which it can be
``pumped" by a laser, and that the atom is practically transparent for
these photons when it is not in this state.  Then, our procedure
selects atoms in the specific state without changing their state in
any way.

It is common to think that unlike classical mechanics, quantum
mechanics poses severe restrictions on the minimal disturbance of the
system due to the measurement procedure.  We, however, have presented
here an ultimately delicate quantum measurement which is impossible to
perform classically.  We found that it is possible to obtain certain
information about a region in space without any interaction in that
region neither in the past nor at present.

{\twelveb ACKNOWLEDGMENTS}

It is a pleasure to thank Shmuel Nussinov, Philip Pearle, and Sandu
Popescu for helpful comments.  The research was supported in part by
grant 425/91-1 of the the Basic Research Foundation (administered by
the Israel Academy of Sciences).

{\twelveb NOTE ADDED IN PROOF}

A few preprints of this paper, written at 1991
(TAUP 1865-91) were circulated among our collegues and gave rise to a
number of publications.$^{12-15}$

\endpage

\bigskip
\tenrm
\tolerance 1500
\hfuzz 6.5pt
{\twelveb REFERENCES}
\smallskip

$^{(a)}$~~Also at the
Department of Chemical Physics,
The Weizmann Institute of Science,
76 100 \ Rehovot, ISRAEL.

1. J. Bell,  {\it Physics}~ {\bf1}, 195 (1964).\hfill\break
2. A. Aspect, J. Dalibard, and G. Roger, {\it Phys. Rev. Lett.}~{\bf
49}, 1804 (1982).\hfill\break
3. Y. Aharonov and D. Bohm, {\it Phys. Rev.}~{\bf 115}, 1804 (1959).
\hfill\break
4. D.M. Greenberger, {\it Rev. Mod. Phys.}~{\bf 55}, 875 (1983).
\hfill\break
5. P. Grangier, A. G. Roger, and A. Aspect, {\it
Europhys. Lett.}~{\bf 1}, 173 (1986).\hfill\break
6. There is also a possibility that the object is transparent and the
photon's wave function have passed through the object while changing
its phase.  The ``click" in $D_2$ tells us that there is an object in
the specified region, but only if later we have found that the object
is not transparent, we can claim that this information was obtained
without interaction.  In Sec. 4 we shall consider a situation in which
we ensure that the measurement is interaction-free by another means.
\hfill\break
7. J. von Neumann, {\it Mathematical Foundations of Quantum Theory
and Measurement}, Princeton, NJ (1983).\hfill\break
8. H. Everett, {\it Rev. Mod. Phys.}~{\bf 29}, 454 (1957).\hfill\break
9. R.H. Dicke,  {\it Am. J. Phys.}~ {\bf 49}, 925 (1981).\hfill\break
10. This experiment gives a clear demonstration of the violation of
conservation laws by the collapse of a quantum state.  For example,
if the potentials in regions A and B are different, then the
expectation value of the particle's energy changes though no change
happens to the photon (see also Ref.9).  Conservation laws are
restored when all branches of the quantum state are considered
together.\hfill\break
11. The MWI presents also a natural explanation why we cannot do
better.
Consider the world in which the photon hits the bomb.  The world which
replaces it in case that the bomb is transparent interfere
destructively with the world in which the detector $D_2$ clicks.
Since the latter is completely eliminated it could not have
probability larger than that of the former.\hfill\break
12. L. Hardy,  {\it Phys. Lett.} {\bf A 160}, 1 (1991).\hfill\break
13. N. Cufaro-Petroni and J.P. Vigier,
 {\it Found. Phys.} {\bf 22}, 1 (1992).\hfill\break
14. L. Hardy,  {\it Phys. Rev. Lett.} {\bf 68}, 2981 (1992).
\hfill\break
15. L. Hardy,  {\it Phys. Lett.} {\bf A 167}, 11 (1992).\hfill\break
\vskip 10 true cm

\noindent
{\bf Fig. 1} Mach-Zehnder type particle interferometer. Detector $D_2$
clicks only if one of the arms of the interferometer is blocked by an
object.

  \end

ct.

  \end

\hangref{%
L. Hardy,  {\it Phys. Rev. Lett.} {\bf 68}, 2981 (1992).}

\hangref{%
L. Hardy,  {\it Phys. Lett.} {\bf A 167}, 11 (1992).}
\vskip 10 true cm

\noindent
{\bf Fig. 1} Mach-Zehnder type particle interferometer. Detector $D_2$
clicks only if one of the arms of the interferometer is blocked by an
object.

  \end